\documentclass[a4paper,singlecolumn,epsfig,showpacs]{revtex4}
\usepackage{amsmath,bm}
\usepackage{epsfig,amssymb}
\usepackage{float}

\begin{document}
\title{A  4-Component Dirac Theory of Ionization of Hydrogen Molecular Ion in a Super-Intense Laser Field} 
\author{
F. H. M. Faisal}
\affiliation
{
Fakult\"at f\"ur Physik, Universit\"at Bielefeld, Postfach 100131, D-33501
Bielefeld, Germany
}
\date{\today}

\begin{abstract}
In this paper a 4-component Dirac theory of ionization of hydrogen molecular ion in a super-intense laser field is presented. Simple analytic expressions for the spin specific as well as the total ionization currents emitted from the ground state of the ion are derived. The results are given for all polarization and finite propagation vectors of the field. They apply for the inner-shell ionization of analogous heavier molecular ions as well. The presence of molecular two-slit interference effect, first found in the non-relativistic case, and the spin-flip ionization current, and an asymmetry of the up- and down-spin currents similar to that predicted in the atomic case, are found also to hold for the present relativistic molecular ionic case. Finally, the possibility of controlling the dominant spin currents by simply selecting the handedness of a circularly polarized incident laser field is 
pointed out.
\end{abstract}
\maketitle

\section{Introduction}
A non-perturbative relativistic analysis of intense-field ionization dynamics becomes necessary if: (a) the laser intensity is so high that the ponderomotive energy $U_{p}= \frac{e_{0}^{2}F^{2}}{4m_{0}\omega^{2}}$
becomes comparable to or greater than the rest-mass energy $m_{0}c^{2}$ of the electron or $U_{p}\ge m_{0}c^{2}$,
(b) the spin degrees of freedom, the magnetic field component and/or the retardation effect  of the field become of interest,
and (c) the initial bound electron is in a deep-lying shell of the target molecule or its ions, where the electron motion is already relativistic. 
It should be also borne in mind that although initially a bound electron may be in the upper valence shell -- where its motion can be non-relativistic ($\beta=\frac{v}{c} << 1$) -- a super-intense laser pulse could, nevertheless,  strip the target molecule of the outer shell electrons before the peak of the pulse arrives. Thus effectively the pulse at its highest intensities might in fact interact with the relativistically moving electrons of the inner-shells of the stripped ions, and ionize them further. Also, an initially non-relativistically moving outer-shell electron on virtual ionization at low velocities might be accelerated to relativistic velocities by the ultra-intense laser pulse. Furthermore, intense {\it high}-frequency free electron laser (FEL) radiations can interact directly and strongly with the relativistically moving bound electrons in the deep-lying shells of atoms and molecules.

In view of the recent progress in ultra-short laser pulse techniques and high-frequency FEL technology, super-intense laser fields reaching intensities $\ge 10^{22}$ W/cm$^{2}$, and FEL radiations of intensities $\ge 10^{16}$W/cm$^{2}$ are now available or expected to be available in the laboratories in the near future. 
The purpose of this paper is to present for the first time a 4-component Dirac theory of ionization of the simplest of molecules, H$_{2}^{+}$, and of the analogous but heavier K-shell molecular ions, by laser fields of arbitrary intensity and polarization and wavenumber.  
The present method of analysis within the strong-field KFR ansatz \cite{KelFaiRei, FaiBha}, is a generalization of the method developed earlier for the atomic case \cite{FaiBha} and can be further generalized to 
the inner-shell ionization of polyatomic molecules in a manner similar to the non-relativistic approach in\cite{MutBecFai,MutBecChiFai}.

\section{Theoretical formulation}

The Dirac Equation of a diatomic molecular-ion interacting with an intense electromagnetic field can be written as:
\begin{eqnarray}\label{DirEqu}
i\hbar \frac{\partial \Psi(t)}{\partial t} &=& [c\bm{\alpha}\cdot (\bm{p}_{op} - \frac{e_{0}}{c}\bm{A}(\omega t-\bm{\kappa}\cdot\bm{r})) - \frac{Z_{a}e_{0}^{2}}{r_{a}} -\frac{Z_{b}e_{0}^{2}}{r_{b}} + \frac{Z_{a}Z_{b}e_{0}^{2}}{R}  + m_{0}c^{2}]\Psi(t) 
\end{eqnarray}
where, $m_{0}$ and $e_{0}$ are the rest mass and charge of electron, 
$\bm{\alpha}$ and $\beta$ are the standard Dirac 4x4 matrices, and $V_{i}(t,\bm{r}) = 
-e_{0}\bm{\alpha}\cdot \bm{A}(u)$, with $u = (\omega t- \bm{\kappa}\cdot\bm{r})$, is the electron-laser interaction Hamiltonian
and $\bm{A}(u)$
is the vector potential of the external radiation field:
\begin{eqnarray}\label{VecPot}
\bm{A}(\omega t- \bm{\kappa}\cdot\bm{r}) &=& A_{0}(\bm{e}_{1} \cos{(\xi/2)} \cos{(\omega t- \bm{\kappa}\cdot\bm{r})} -\bm{e}_{2} \sin{(\xi/2)} \sin{(\omega t- \bm{\kappa}\cdot\bm{r})})
\end{eqnarray}
We consider the general case of a laser field of arbitrary elliptic polarization 
(ellipticity parameter $\xi [0,\pm \pi/2]$; $\pm$ stand for the left and right helicity) with the semi-major and the semi-minor 
unit polarization vectors $\bm{e}_{1}$ and $\bm{e}_{2}$, respectively.
It should be noted that the Coulomb fields of the ``heavy'' nuclei a and b are treated as {\it external} fields acting on the electron at the position $\bm{r}$. We have defined, $r_{a }= |\bm{r}-\bm{R}_{a}| $ and $r_{b }= |\bm{r}-\bm{R}_{b}| $; $\bm{R}_{a}$, $\bm{R}_{b}$ are the locations of the two nuclei a and b, respectively. Clearly, for the homonuclear diatomic case (C.M. of the nuclei as coordinate center),
$\bm{R}_{a}=-\bm{R}_{b} = \bm{R}/2$ where $R$ is the nuclear separation, and $Z_{a}=Z_{b}=Z$ is the nuclear charge.  

We introduce the initial state partition of the total Dirac Hamiltonian
 \begin{equation}\label{VolEqu}
H_{D}(t) = H_{mol} + V_{i}(u)
\end{equation}
where, $V_{i}(u)=-e_{0}\bm{\alpha}\cdot\bm{A}(u)$, and
\begin{equation}\label{VolEqu}
H_{mol} = [c\bm{\alpha}\cdot \bm{p}_{op} - 
\frac{Z_{a}e_{0}^{2}}{r_{a}} -\frac{Z_{b}e_{0}^{2}}{r_{b}} + \frac{Z_{a}Z_{b}e_{0}^{2}}{R}  
+ m_{0}c^{2}] 
\end{equation}
is the molecular Dirac Hamiltonian, with stationary eigenfunctions satisfying the Dirac equation:
\begin{equation}\label{VolEqu}
(E_{i} - H_{mol})\psi_{i} (\bm{r}) = 0 
\end{equation} 
with positive and/or negative energy eigenvalues $E_{i}$.
We choose the final state partition of the total Dirac Hamiltonian $H_{D}(t)$ (given by the right hand side of Eq. (\ref{DirEqu})) as:
$H_{D}(t) = H_{Vol}(t) + V_{f}$
where,   
\begin{equation}\label{DirVolEqu}
H_{Vol}(t) = (c\bm{\alpha}\cdot (\bm{p}_{op} - \frac{e_{0}}{c}\bm{A}(u)) + m_{0}c^{2})
\end{equation}
is the Dirac-Volkov Hamiltonian of an electron in the plane wave field, Eq. (\ref{VecPot}), and 
\begin{equation}\label{FinInt}
V_{f}=V_{mol}(\bm{r};R)=-\frac{Z_{a}e_{0}^{2}}{r_{a}} -\frac{Z_{b}e_{0}^{2}}{r_{b}} + \frac{Z_{a}Z_{b}e_{0}^{2}}{R}.  
\end{equation}
For ease of working with Dirac equations it is useful to reduce the number of 4 natural constants appearing in them to only two by introducing 
the reduced constants: $m=\frac{m_{0}c}{\hbar}$ and $e=\frac{e_{0}}{\hbar c}$
(which can be therefore restored easily in the end, if desired),  
and the following Feynman notations for 4-vector, 4-scalar product and the 4-gamma matrices: \\
%\begin{eqnarray}\label{FenNot}
\begin{eqnarray}\label{FenNot}
\kappa_{0}&=&\omega/c\nonumber\\
\kappa &=& (\kappa_{0}, \bm{\kappa})\nonumber\\ 
a &=& (a_{0}, \bm{a})\nonumber\\
a\cdot b &=& a_{0}b_{0} -\bm{a}\cdot\bm{b}\nonumber\\ 
u&=&\kappa\cdot x\nonumber\\
\gamma_{0} &=& \beta \nonumber\\
\bm{\gamma} &=& \beta\bm{\alpha}\nonumber\\
\not\!a &=& \gamma \cdot a\nonumber\\
\partial &=&(\partial_{0} ,\partial_{\bm{x}})\nonumber\\
\bar{\phi} &=& \phi^{\dagger}\gamma_{0}\nonumber\\
%&=& \frac{\partial}{\partial x}\nonumber\\
d^{4}x &=& dx_{0} d\bm{x}.
\end{eqnarray}
In this notation, for example, Eq. (\ref{DirEqu}) takes the form:
\begin{equation}\label{DirEquTwo}
(i\not\!\partial - e\not\!\!A(x) - \gamma_{0}V_{mol}(x;R) -m)\Psi(x)=0.
\end{equation}
The total Dirac wavefunction can be expanded systematically, as in the non-relativistic ``intense-field many-body S-matrix theory'' (or IMST, review \cite{BecFai}), in terms of the Dirac-Volkov-Feynman Green's function $G_{F}(x,x')$ 
that satisfies:
\begin{equation}\label{VolEqu}
(i\not\!\partial - e\not\!\! A(u) - m)G_{F}(x,x') = \delta^{4}(x-x'). 
\end{equation} 
%(i\gamma\cdot\partial - e\gamma\cdot A(u) - m)G_{f}(x,x') = \delta^{4}(x-x') 
A complete set of solution \cite{Vol} of the Dirac-Volkov equation (\ref{DirVolEqu}) can be conveniently written as:
\begin{equation}\label{VolWavFun}
\Phi_{p}^{(s)}(x) = \sqrt{\frac{m}{p_{0}}}e^{-ip\cdot x - if_{p}(x)} (1+ \frac{e\not\!\!\kappa\not\!\! A(u)}
{2p\cdot\kappa}) u^{(s)}_{p}
\end{equation}
where, for a general pulse of transverse external electromagnetic wave, $A(u)$, we get: 
\begin{equation}
f_{p}(x) = \int_{0}^{\kappa\cdot x}(\frac{e A(u)\cdot p}{\kappa\cdot p} -\frac{e^{2}A^{2}(u)}{2\kappa\cdot p}) du . 
\end{equation}
For a vector potential of a constant envelope it reduces to:
$f_{p}(x) = -a_{p} \sin{(u+\chi_{p})} + b_{p} \sin{(2u)} + \zeta_{p} u$,
with,
$a_{p}=\frac{A_0\left|\vec{\epsilon}(\xi)\cdot\vec{p}\right|}{c\kappa\cdot p}$,
$b_{p}=\frac{A_0^2}{8c^2 \kappa\cdot p} \cos \xi$,
$\chi_{p}=\tan^{-1}[\tan\phi_p \tan(\xi/2)]$; $u_{p}^{(s)}$ are the four (two positive and two negative energy) Dirac plane wave spinors:
\begin{eqnarray}
u^{(s)}(p) &=&\frac{\not\!p +m}{\sqrt(2m(m+E)}{w}^{(s)}, s=(1,2), (p_{0} = E) \\ 
u^{(s)}(-p)&=& \frac{-\not\!p +m}{\sqrt(2m(m+E)}{w}^{(s)}, s=(3,4), (p_{0}= -E) 
\end{eqnarray}
where, $E \equiv +\sqrt{(\bm{p}^{2}+m^{2})}$ is always positive.
The up and down positive energy 4-spinors ${w}^{(s)}$ are given by: 
\begin{subequations}\label{eq:bsomega}
\begin{equation}
{w}^{(1=u)} = \begin{pmatrix}
                    1\\
                    0\\
                    0\\
                    0
                  \end{pmatrix}\label{eq:omegaup}
\end{equation} and
\begin{equation}
{w}^{(2=d)} = \begin{pmatrix}
                      0\\
                      1\\
                      0\\
                      0
                    \end{pmatrix}\label{eq:omegadp}\; .
\end{equation}
\end{subequations} 
Similarly, the up and down negative energy 4-spinors are:
\begin{subequations}\label{eq:bsomega}
\begin{equation}
{w}^{(3=u)} = \begin{pmatrix}
                    0\\
                    0\\
                    1\\
                    0
                  \end{pmatrix}\label{eq:omegaum}
\end{equation} and
\begin{equation}
{w}^{(4=d)} = \begin{pmatrix}
                      0\\
                      0\\
                      0\\
                      1
                    \end{pmatrix}\label{eq:omegadm}\; .
\end{equation}
\end{subequations}

Thus, the Dirac-Vokov Green's function satisfying the Feynman-St\"uckelberg  boundary condition (cf. \cite{Fey}), can be defined by: 
\begin{eqnarray}\label{VolGreFun}
G_{F} (x,x') &=& -i(\frac{m}{E})\sum_{\bm{p}} (\sum_{s=1,2}
\theta{(t-t')} \Phi^{(s)}_{p}(x){\bar{\Phi}}^{(s)}_{p}(x') \nonumber\\
&& + \sum_{s=3,4}\theta{(t'-t)} \Phi^{(s)}_{-p}(x){\bar{\Phi}}^{(s)}_{-p}(x') )
\end{eqnarray}
where, $\sum_{\bm{p}} =\frac{1}{(2\pi)^{3}}\int d^{3}p$.
For an explicit momentum representation of $G_{F}(x,x')$, which can be used conveniently for intense periodic 
electromagnetic fields, we refer to \cite{FaiVolGreFun}, or Eq. (54) of \cite{BecFai}. 

Using $G_{F}(x,x')$ one may systematically solve Eq.(\ref{DirEquTwo}) by iteration, to obtain the total Dirac state evolving from a specified initial state $\psi^{(s)}_{i}(x)$:
\begin{eqnarray}\label{TotDirWavFun}
 \Psi_{i}(x) &=& \psi^{(s)}_{i}(x)  + \int_{x_{i}}^{x_{f}} d^{4}x' G_{F}(x,x') (e\not\!\! A(x'))\psi^{(s)}_{i}(x') \nonumber\\
 & & + \int_{x_{i}}^{x_{f}} d^{4}x'' \int_{x_{i}}^{x_{f}} d^{4}x' G_{F}(x,x'') (\gamma_{0} V_{f}(x''))\nonumber\\
 & & \times G_{F}(x'',x') (e\not\!\! A(x'))\psi^{(s)}_{i}(x') + \cdots\cdots.
\end{eqnarray} where $V_{f}(x)$ is the external rest potential. 
This constitutes the Dirac wavefunction within the ``relativistic intense-field many-body S-matrix theory'' (or RIMST) 
subjected to the Feynman-St\"uckelberg  backward time positron interpretation of the negative energy ``Dirac sea''.
It should be noted therefore that the present approach can be used for obtaining the probability of not only ionization, as we consider in this work, but also of such processes as pair creation in super-intense laser fields (cf. \cite{Fey}).

\section{Ground State H$_{2}^{+}$ Molecular Ion in a Dirac Basis}. 

We shall represent the ground state four-component wavefunction $\psi_{i}^{(s)}(x)$, of hydrogenic molecular ions, in the spin state, $s=(u,d)$, by a linear combination of relativistic atomic basis orbitals of $\sigma$ symmetry. 
To this end we define a four-component  ``Dirac basis'' (with $\sigma$ symmetry) of the form:
\begin{eqnarray}\label{DirBas} 
\phi_{j}^{(s)}(\bm{r}) &=& R_{j}(r)\not\!\!n_{j}({\hat{r}}){w}^{(s)}; j=1,2,3\cdots.\\
R_{j}(r) &=& N_{j} r^{\nu_{j}-1}e^{-\lambda_{j} r}
\end{eqnarray}
\noindent where \qquad $\not\!\!\!n_{j}(\hat{r})=\gamma\cdot n_{j}(\hat{r})$, and 
the 4-vector $n_{j}(\hat{r}) = (n_{0}, \bm{n}_{j}(\hat{r}))$, with $n_{0}=1$, and $ \bm{n}_{j}(\hat{r}) = i \beta'_{j} \hat{r}$, and 
$\beta'_{j} =\frac{\lambda_{j}}{m(1+\nu_{j})}$; 
where, \noindent $\nu_{j}$ and $\lambda_{j}$ are in general {\it non}-integer real parameters and $N_{j}= (2\lambda_{j})^{\nu_{j}+1} (\frac{1+ \nu_{j}}{8\pi\Gamma(1+2\nu_{j})})^{1/2} $ is a normalization constant.  
Note that the well known hydrogenic Dirac s-state wavefunctions \cite{BeoDre}
in the present notation (from \cite{Fai} p. 418) can be reproduced by choosing a {\it single} Dirac basis state with
$\nu_{1} = \sqrt{1-(Z\alpha)^{2}}$ and $\lambda_{1} = mZ\alpha$, where $\alpha$ is the fine structure constant and $Z$ is the nuclear charge. 
The Fourier transform of  the Dirac basis functions introduced above are evaluated analytically and we get: 
\begin{eqnarray}\label{FouTra}
\tilde{\phi}_{j}^{(s)} (\bm{q}) &=& \int d^{3}r e^{-i\bm{q}\cdot\bm{r}}\phi_{j}^{(s)}(\bm{r}) = N_{j} c_{j}^{0}(q)[1+ g_{j}(q) (\bm{\gamma}\cdot\hat{\bm{q}})] {w}^{(s)}
\end{eqnarray}
where, 
\begin{eqnarray}\label{eq:c0q}
c_{j}^{0}(q) & = &  \int e^{-i\vec{q}\cdot\vec{r}}r^{\nu_{j}-1}e^{-\lambda_j r}d^3 r \nonumber\\
 &=&   \frac{4\pi\Gamma(1+\nu_{j})}{q(q^2+\lambda_{j}^2)^{\frac{\nu_{j}+1}{2}}} s^{(0)}_{j}(q),  
\end{eqnarray}
 \begin{equation}
    \label{eq:s0q}
    s_{j}^{0}(q)= \sin{\left((\nu_{j}+1)\arctan{(q/\lambda_{j})}\right)}\;,
  \end{equation}
and,
\begin{eqnarray}
  \label{eq:c1q}
  g_{j}(q)  &=& \beta'_{j} [ \frac{\lambda_j}{q} - 
  \frac{\nu_j+1}{\nu_j}\sqrt{1+(\lambda_{j}/q)^2} \frac{\sin{(\nu_j \arctan{(q/\lambda_j)})}}
  {\sin{((\nu_j+1)\arctan{(q/\lambda_j)})} } ].
\end{eqnarray}
The corresponding negative energy basis functions can be obtained from the F.T. (momentum representation) of the above positive energy basis functions by letting $q \rightarrow - q$, and replacing the positive energy up and down spinors 
${w}^{(s)}, s=(1,2)$, by the negative energy up and down spinors ${w}^{(s)}, s=(3,4)$ defined above.
Finally, we write the Dirac H$_{2}^{+}$ ground state wavefunction as:
\begin{equation}\label{GroStaWavFun}
\psi_{i}^{(s)}(\bm{r};R) = \sum_{j} a_{j} [ R_{j}(r_{a}) \not\!\!n_{j}(\hat{r}_{a}){w}^{(s)} + R_{j}(r_{b}) \not\!\!n_{j}(\hat{r}_{b}){w}^{(s)} ]
\end{equation}
where, $a_{j}$ are numerical constants, and the ${w}^{(s)}$ are the positive energy spinors defined above.

\section{Ionization Transition Amplitude: Up and Down Spin Currents}

Projecting the final Volkov state of 4-momentum $p$, Eq. (\ref{VolWavFun}), on the total wavefunction $\Psi_{i}(x)$, and retaining the leading significant term, we get the individual spin selected ionization amplitudes ($s=u$ or $d$ 
to $ s'= u$ or $d$): 
\begin{eqnarray}
  \label{eq:selampion}
  {\cal{A}}_{ion}(s',s) & = & -i\int d^{4}x \bar{\Phi}_{p}^{(s')}(x)e\not\!\!A(x)\psi_{i}^{(s)}(x)
  \end{eqnarray}
Using the Fourier transform of the ground state wavefunction, 
carrying out the $d^{4}x$ integration, and performing the somewhat lengthy Dirac $\gamma$-algebra, 
we arrive at the following {\it explicit} analytic result (in a.u. $e_{0}=m_{0}=\hbar=\alpha c=1$):
\begin{equation}\label{Amp2}
{\cal{A}}_{ion} (s\rightarrow s') = 2\pi i \sum_{n} \delta (p_{0} -p_{b} + (\zeta_{p} -n)\kappa_{0}) T^{(n)}(s',s)
 \end{equation}
with,  
 \begin{eqnarray}\label{Amp2}
T^{(n)}(s',s) &=& \frac{A_{0}}{2c} N_{p_{0}}(e^{i\bm{q}\cdot\bm{R}_{a}} + e^{i\bm{q}\cdot\bm{R}_{b}}) \times t^{(n)}(s',s)\\
t^{(n)}(s',s) &=& \sum_{j} a_{j} N_{j} c_{j}^{(0)}(q) M_{j}^{(n)}(s',s)
 \end{eqnarray}
Given the matrix elements $M_{j}^{(n)}(s',s)$,
the spin-specific {\it rates} of ionization become, 
\begin{eqnarray}\label{Ratss'}
\frac{dW_{s\rightarrow s'}}{d\Omega} 
&=&\sum_{n\geq n_0}\left(\frac{A_{0}}{2c}N_{p_0}\right)^{2} 
\left| T^{(n)}(s',s)\right|^{2} \frac{cp_0|\bm{p}|}{(2\pi)^2}\nonumber\\
&=&(\frac{A_{0}}{2c} N_{p_{0}})^{2}\left|(e^{i\bm{q}\cdot\bm{R}_{a}} + e^{i\bm{q}\cdot\bm{R}_{b}})\right|^{2} \times \left|t^{(n)}(s',s)\right|^{2} \frac{cp_0|\bm{p}|}{(2\pi)^2}\nonumber\\
\end{eqnarray}
The number of absorbed photons (denoted by $n$ above, where $n_{0}$ is the threshold number) is determined by the
4-momentum conservation relation, obtained from the delta function that appears naturally in Eq. (\ref{Amp2}) from the space-time integration: 
$ n\omega = \epsilon_b + \epsilon_{kin} + \zeta_p\omega$,
where $\epsilon_b= c\left(c - \sqrt{c^2- p_b^2}\right)$ is the binding energy
(ionization potential) and $\epsilon_{kin}= c\left(\sqrt{c^2 +p^2}
-c\right)$ is the kinetic energy. 
On completing the Dirac algebra, the reduced $M_{j}^{(n)}(s', s)$-matrix elements are found to be given by the following
simple algebraic expressions: 
\begin{eqnarray}
 \label{eq:An1}  
M_{j}^{(n)} (u \rightarrow u)& = & {B_n^0}^* \left(m_1 + m_2 g_{j}(q)
                                ({\hat{p}}\cdot{\hat{q}} + i(\hat{\bm{p}}\times \hat{\bm{q}})_{z})\right)
                            +\bm{B}_n^*\cdot\left(m_2{\hat{p}} +m_1g_{j}(q){\hat{q}}\right)\nonumber\\
                           &   & -i\left(\bm{B}_n^*\times\left(m_2
                                     {\hat{p}} - m_1g_{j}(q){\hat{q}}\right)
                                   \right)_{z},
\end{eqnarray}
\begin{eqnarray}
  \label{eq:An2}
  M_{j}^{(n)}(u\rightarrow d)& = & m_2g_{j}(q){B_n^0}^*
                                 \left(i\left({\hat{p}}
                                     \times{\hat{q}}\right)_{x}
                                  - \left({\hat{p}}
                                     \times {\hat{q}}\right)_{y} \right)
                            -i\left(\bm{B}_n^*\times\left(m_2
                                     {\hat{p}}-m_1g_{j}(q){\hat{q}}\right)
                                   \right)_{x}\nonumber\\
                           &   & +\left(\bm{B}_n^*\times\left(m_2
                                     {\hat{p}}-m_1g_{j}(q){\hat{q}}\right)
                                   \right)_{y},
\end{eqnarray}
\begin{eqnarray}
  \label{eq:An3}
  M_{j}^{(n)} (d\rightarrow u)& = & m_2g_{j}(q){B_n^0}^*
                                 \left(i\left({\hat{p}}
                                     \times{\hat{q}}\right)_{x}
                                  + \left({\hat{p}}
                                     \times{\hat{q}}\right)_{y}
                                 \right)
                           -i\left(\vec{B}_n^*\times\left(m_2
                                     {\hat{p}}-m_1g_{j}(q){\hat{q}}\right)
                                   \right)_{x}\nonumber\\
                           &   & -\left(\vec{B}_n^*\times\left(m_2
                                     {\hat{p}}-m_1g_{j}(q){\hat{q}}\right)
                                   \right)_{y},
\end{eqnarray}
\begin{eqnarray}
  \label{eq:An4}
  M_{j}^{(n)} (d\rightarrow d) & = & {B_n^0}^* \left(m_1 + m_2 g_{j}(q)(
                               {\hat{p}}\cdot{\hat{q}} -   i(\hat{\bm{p}}\times \hat{\bm{q}})_{z}) \right)
                            +\bm{B}_n^*\cdot\left(m_2{\hat{p}}
                                   +m_1g_{j}(q){\hat{q}}\right)\nonumber\\
                           &   & +i\left(\bm{B}_n^*\times\left(m_2
                                     {\hat{p}}-m_1g_{j}(q){\hat{q}}\right)
                                   \right)_{z}.
\end{eqnarray}\\
where,
\begin{eqnarray}
B_n^0 & = &  \frac{A_0 \kappa_0}{4c{\kappa}\cdot{p}}
  \left(2 {\cal{J}}_n + \cos\xi\left({\cal{J}}_{n+2} + 
  {\cal{J}}_{n-2} \right)\right)\;,\\
\bm{B_n} & = &  \bm{e}(\xi) {\cal{J}}_{n-1} 
 + \bm{e}^{*}(\xi){\cal{J}}_{n+1}
 + {\hat{\kappa}}B_n^0\;,\\
\kappa\cdot p & = & \kappa_0 k_0 - \bm{\kappa}\cdot\bm{p}, \kappa=(\kappa_0,\bm{\kappa}), \kappa_0  =  \omega/c\;.
\end{eqnarray}
\begin{equation}
 {\cal{J}}_n \equiv  {\cal{J}}_n(a_{p},b_{p},\chi_{p}) =
\sum_m J_{n+2m}(a_{p})J_m(b_{p})e^{i(n+2m)\chi_{p}}
\end{equation}
are generalized Bessel functions of three arguments.
The other parameters are the ``field-dressed'' electron momentum
$\bm{q}\equiv q {\hat{q}}
          = \bm{p} + \left(\zeta_{p}-n\right)\bm{\kappa}$,
$\zeta_{p}=\frac{A^2}{4c^2 \kappa\cdot p}$,
${\bm{e}}(\xi)=[\bm{e}_1\cos(\frac{\xi}{2}) + i \bm{e}_2\sin(\frac{\xi}{2})]$, with $\xi= [0,\pm\pi/2]$,
$p_0=\sqrt{c^2 + \bm{p}^2}= \left(n-\zeta_{p}\right)\kappa_0
+ \sqrt{c^2 - p_b^2}$,
$ m_1      =    \sqrt{(p_0 + c)/(2c)}$, 
$ m_2      =  - \sqrt{(p_0 - c)/(2c)}$, 
and $ N_{p_0}  =  \sqrt{\frac{c}{p_0}}$.

Finally, the spin {\it unresolved} total ionization rate
from an {\it unpolarized} target atom is
obtained by simply adding the four spin-specific
rates given above and dividing by $2$ (for the average with respect to
the two degenerate initial spin states):
\begin{eqnarray} \label{TotRat}
\frac{d\Gamma^{(+)}}{d\Omega} &=&\frac{1}{2}\sum_{n\geq n_0}\left(\frac{A_{0}}{2c}N_{p_0}\right)^{2} \nonumber\\
& & \times |e^{i\bm{q}\cdot\bm{R}_{a}} + e^{i\bm{q}\cdot\bm{R}_{b}}|^{2}\nonumber\\
&&\times \sum_{(s',s)=(u,d)} \left|\sum_{j} a_{j}N_{j} c_{j}^{(0)}(q)  M_{j}^{(n)}(s',s)\right|^2 \nonumber\\
& & \times c p_0\frac{|\bm{p}|}{(2\pi)^2} 
 \end{eqnarray}
We may point out that Eqs. (\ref{Amp2}), (\ref{Ratss'}) and (\ref{TotRat}) that apply directly for the $\sigma_{g}$ case, can be used to obtain also the probability of ionization from an analogous $\sigma_{u}$ state, if one simply replaces the sum  (``+'') of the two interfering nuclear terms in these formulas with their difference (``-'').

\subsection{Factorization of ionization probability and molecular two-slit interference} 

It is interesting to see from Eqs. (\ref{Amp2}), (\ref{Ratss'}) and (\ref{TotRat}) that the ionization amplitude and the probabilities factorize essentially 
into two factors, one of which depends on the nuclear geometry and the other on the electronic degrees of freedom. 
The geometrical factor leads to an interference effect arising from the superposition of the partial electronic waves that emerge from the two nuclear centers with a phase difference, $\bm{q}\cdot(\bm{R}_{b} - \bm{R}_{a}) = \bm{q}\cdot\bm{R}$. 
This leads to the now well-known molecular two-slit interference (and its dependence on the symmetry 
of the electronic wavefunction of the active electron) first discussed in the non-relativistic case \cite{MutBecFai}.  
Thus for example, for the ionization probability from the   
$\sigma_{g}$ molecular orbital of H$_{2}^{+}$, the two-slit interference factor becomes:
$ | 1+ e^{i\bm{q}\cdot\bm{R}}|^{2}$ which for $\bm{q}\cdot\bm{R} = 2n\pi, n=0,1,2,...$  gives rise to a constructive interference  and an
enhancement  of the ionization probability by a maximum factor of  4, while, for $\bm{q}\cdot\bm{R} = (2n+1)\pi, n=0,1,2,...$,
a destructive interference yields a minimum in the energy-momentum distributions of the ionized electron.
We note that the two-slit interference effect is {\it generic} in nature 
and can show its effect in all channels in which the active electron of the molecule determines the channel probability, not only in ionization but also e.g. in laser stimulated molecular electron diffraction, laser induced emission of (low or high) harmonic radiation etc. Furthermore,  it can survive even in the signals that 
are obtained in orientation averaged measurements.
In this case the interference probability factor reduces to:
 \begin{equation}\label{AveIntFac}
 \frac{1}{4\pi}\int d\Omega_{R} |1+e^{i\bm{q}\cdot\bm{R}}|^{2}= 2(1+ \frac{\sin{qR}}{qR})                    
\end{equation} 
which indicates an interference modulation with the variation of the electron momentum q (or R), albeit decreasing in the modulation-depth with increasing momentum and/or the nuclear separation R.
 
\subsection{Linear and Circular Polarization, Photon Helicity and Spin Asymmetry}

The results derived above hold for any polarization of the laser field. We may note below the particular simplifications that arise in the two most common cases of linear and circular polarization.

1. Linear polarization:\\
In the case of a linearly polarized laser field, the explicit formulas for the ionization currents,
Eqs. (\ref{Ratss'}) and (\ref{TotRat}), hold again when we simply put  
$\xi = 0$, in them. Thus the only change occurs in replacing the generalized Bessel function of three arguments
by the generalized Bessel function of two arguments: ${\cal{J}}_n = J_n(a_{p},b_{p})$, with $a_{p}=a_{p}(\xi=0), b_{p}=b_{p}(\xi=0)$.\\
 
 2. Circular polarization:\\
Similarly, for the case of circular polarization of the laser field, the same expressions hold again when we simply put
$\xi= \pm \pi/2$ in them. In this case the generalized Bessel function of three arguments reduces to an ordinary Bessel function (of one argument) times a simple phase factor: ${\cal{J}}_n = J_n(a_{p})e^{\pm i n\phi_p}$, with $a_{p} = a_{p}(\xi=\pm\pi/2)$, where $\pm$ stand for the positive and negative handedness of the circularly polarized laser field. \\
Closer examinations of the expressions of the matrix elements $M_{j}^{(n)}(s',s)$  for circular polarization, as in the atomic case \cite{FaiBha}, 
show that the probability of spin-flip ionization for $u\rightarrow d$ 
transition {\it differ} from the probability of the $d\rightarrow u$ transition. 
%This situation is reminiscent of the corresponding problem of atomic ionization in (cf. \cite{FaiBha}).  
It also suggests that the up and down spin-currents, even from a spin {\it unpolarized} 
(or equal mixture of up and down spin) ground state would show an {\it asymmetry} in any direction of emission of the electrons. This asymmetry is expected to persists even when the laser photons are of long-wavelength
(or, the laser propagation vector $\kappa \rightarrow 0$). 
This implies that the dominant 
component of the electron spin-current emitted from molecular targets by intense laser fields might be {\it controlled} by choosing the helicity of an incident circularly polarized laser field. 

\section{Summary}
In this paper, a 4-component Dirac relativistic theory of interaction of super-intense laser fields with H$_{2}^{+}$ 
molecular ion, and its heavier K-shell analogs, is presented.  To this end a ``Dirac basis'' of atomic orbitals is introduced and employed to construct the LCAOMO of the ground state of H$_{2}^{+}$. 
%It also applies by analogy to the K-shell molecular ions of $\sigma_{g,u}$ symmetry. 
Explicit analytic expressions for the probability of ionization per second by an incident laser field of arbitrary intensity, wavelength and polarization is derived. 
The formulas derived apply for the total ionization probability as well as for the spin specific measurments of the ionization signal. They
imply, among other things, an asymmetry of the probability of spin-flip, and of the up-spin and down-spin ionization currents, even from a spin {\it unpolarized} ground state molecular ion.
The presence of a {\it generic} molecular two-slit interference effect, first noted in the non-relativistic case \cite{MutBecFai}, is further confirmed in the present relativistic analysis.  Finally, the possibility of controlling the dominant spin component of the electron current emitted from the inner-shell of a molecule by simply selecting the handedness of a circularly polarized incident laser field is pointed out. 

%\end{subequations}
\end{document}